\newcommand\apjl{{ApJ\,}}%
\newcommand\apjs{{ApJS\,}}%
\newcommand\apss{{Ap\&SS\,}}%
\newcommand\aap{{A\&A\,}}%
\newcommand\mnras{{MNRAS\,}}%
\newcommand{\Kepler}{\textit{Kepler}}

\documentclass[graybox, natbib, footinfo]{svmult}

\usepackage{url}

\usepackage{mathptmx}       
\usepackage{helvet}         
\usepackage{courier}        
\usepackage{type1cm}        

\usepackage{makeidx}         
\usepackage{graphicx}        
\usepackage{multicol}        
\usepackage[bottom]{footmisc}

\makeindex             

\begin{document}

\title*{Impact of rotational mixing on the global and asteroseismic properties of red giants}
\author{Patrick Eggenberger, Nad{\`e}ge Lagarde and Corinne Charbonnel}
\institute{P. Eggenberger \at Observatoire de Gen\`eve, Universit\'e de Gen\`eve, 51 ch. des Maillettes,
CH-1290 Sauverny\\ \email{patrick.eggenberger@unige.ch}
\and N. Lagarde \at Observatoire de Gen\`eve, Universit\'e de Gen\`eve, 51 ch. des Maillettes,
CH-1290 Sauverny\\ \email{nadege.lagarde@unige.ch}
\and C. Charbonnel \at Observatoire de Gen\`eve, Universit\'e de Gen\`eve, 51 ch. des Maillettes,
CH-1290 Sauverny \\ Laboratoire d'Astrophysique de Toulouse-Tarbes, CNRS UMR 5572, Universit\'e de Toulouse, 14 av. E. Belin F-31400 Toulouse\\ \email{corinne.charbonnel@unige.ch}
}
%
%
\maketitle

\abstract{
The influence of rotational mixing on the global parameters and asteroseismic properties of red giant stars 
is reviewed. While red giants are generally characterised by low surface rotational velocities, they may have been rotating much more rapidly during the main sequence, so that the rotational history of a star has a large impact on its properties during the red giant phase. 
For stars massive enough to ignite helium burning in non-degenerate conditions, rotational mixing leads to a significant increase of the stellar 
luminosity and shifts the location of the core helium burning phase to a higher luminosity in the HR diagram. This results in a change 
of the seismic properties of red giants and of the fundamental parameters of a red giant star as determined by performing 
an asteroseismic calibration. For red giants with a lower mass that undergo the helium flash, rotational mixing decreases the luminosity
of the bump at solar metallicity changing thereby the global and asteroseismic properties of these stars. }

\section{Introduction}
\label{sec:intro}

The wealth of information obtained about the internal structure of the Sun by the observation and interpretation of solar oscillation modes 
stimulated various attempts to detect solar-like oscillations on other stars. Consequently, many instruments have been recently developed to perform such asteroseismic studies. From the ground, the stabilized spectrographs have reached the accuracy needed to detect and characterise solar-like oscillations on stars other than the Sun, while space missions have been recently designed to perform very high accuracy photometric measurements of these oscillation modes. Solar-like oscillations are not restricted to solar-type stars, but are expected in any star exhibiting a convective envelope able to stochastically excite pressure modes of oscillations. In the case of red giants, solar-like oscillations have been first detected for a few stars \citep[see e.g.][]{fra02, bar04, der06, bar07},  while clear identifications of non-radial oscillations have then been obtained for a large number of red giant stars with the CoRoT space mission \citep{der09} and the \Kepler\ satellite \citep{bed10}.  These observations stimulated population studies aiming at reproducing the distribution of global asteroseismic properties of red giant stars \citep{mig09a,mig09b}, as well as the theoretical study of the asteroseismic properties of red giants and of 
the effects of various physical processes on the modelling of these stars \citep{dup09,egg10_rg,mon10}. 
Rotation being one of the key processes that has an important impact on stellar physics and evolution \citep{mae09}, we briefly review here its effects on the global and asteroseismic properties of red giants.

\section{Effects of rotation on evolutionary tracks}
\label{sec_dhr}

Stellar models computed with and without the inclusion of shellular rotation \citep{zah92} are compared in order to discuss the influence of rotation on the evolution and global properties of red giant stars. We discuss first the evolution of stars massive enough to start helium burning in non-degenerate conditions, and then the evolution of low mass red giants undergoing the helium flash. 

\subsection{Intermediate-mass stars}
\label{sec_dhr_m3}

\begin{figure}[htb!]
\includegraphics[scale=.55]{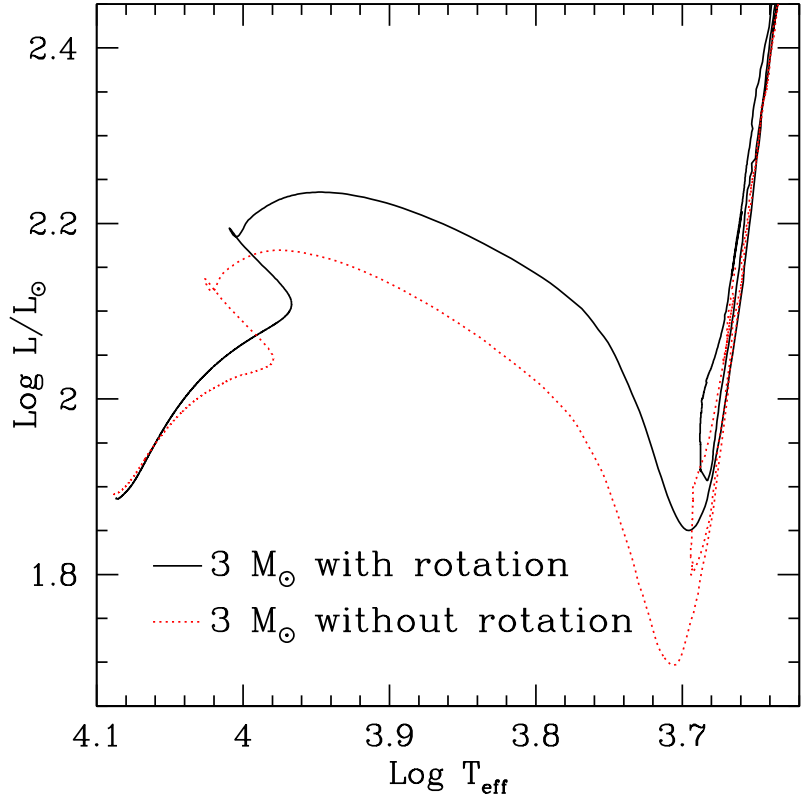}
\caption{Evolutionary tracks in the HR diagram for 3\,M$_\odot$ models computed with (continuous line) and without (dotted line) shellular rotation.}
\label{dhr_m3}
\end{figure}

The effects of rotation on the global properties of red giants massive enough to ignite He burning in non-degenerate conditions are illustrated by computing the evolution of a 
3\,M$_{\odot}$ star with and without rotation using the Geneva stellar evolution code \citep[][]{egg08}.
The solar chemical composition given by \cite{gre93} is used together with a 
solar calibrated value for the mixing-length parameter. No overshooting from the convective core into the surrounding radiatively 
stable layers is included. Figure~\ref{dhr_m3} compares the evolutionary track of the non-rotating model with that of a rotating model computed with exactly the same input parameters except for an initial velocity of 150\,km\,s$^{-1}$ on the zero age main sequence (ZAMS). 

We see that evolutionary tracks in the HR diagram are significantly affected when rotational effects are taken into account. 
During the main sequence, we first note a slight decrease of the luminosity for the rotating model compared to the model without rotation at the beginning of the main-sequence evolution, while the increase of the luminosity of the rotating model becomes more and more pronounced as evolution proceeds. 
In particular, the model including rotation has a larger luminosity than the standard one during the post-main sequence phase of evolution. The location of the core helium-burning phase, which is the phase during which the star spends most of its lifetime
as a red giant, is then shifted to larger luminosity when rotational effects are included. 

We recall here that rotational effects can be basically divided in two main categories: (1) the effects resulting from changes in the chemical and internal structure induced by rotational mixing and (2) the effects resulting from hydrostatic corrections due to the centrifugal force. 
At the beginning of the main sequence, only hydrostatic effects of rotation are observed. The effective gravity of the rotating model is then slightly reduced compared to the non-rotating model so that the resulting evolutionary track is similar to the one of a non-rotating star computed with a slightly lower initial mass. This explains the small decrease in luminosity and effective temperature observed near the ZAMS for the model including rotation in Fig.~\ref{dhr_m3}. As the star evolves on the main sequence, the effects related to rotational mixing plays however a more and more important role by bringing fresh hydrogen fuel in the stellar core and transporting helium and other H-burning products in the radiative zone. Rotational mixing thus increases the size of the convective core and changes the chemical composition profiles in the radiative zone. This induces an increase of the luminosity together with a widening of the main sequence when rotation is included in the computation (see Fig.~\ref{dhr_m3}). The changes observed in the tracks are thus mainly due to rotational mixing with only a very limited contribution from the effects of the centrifugal force at the very beginning of the main sequence. This can be explained by recalling that the kinetic rotational energy of the 3\,M$_\odot$ star is much lower than its gravitational energy.

Rotational mixing has also an important influence on stellar ages. The value of the central abundance of hydrogen at a given age is indeed larger for the rotating model than for the non-rotating one. This is due to the transport of hydrogen  
in the deep stellar layers, which leads to an enhancement of the main-sequence lifetime for rotating models compared to standard models. For the 3\,M$_\odot$ model considered here with an initial velocity of 150\,km\,s$^{-1}$, this corresponds to an increase of the age of about 10\,\%. 

As far as the evolution of the surface rotational velocity is concerned, a slow decrease is first obtained during the main sequence; starting with an initial velocity of 150\,km\,s$^{-1}$ on the ZAMS, a velocity of 115\,km\,s$^{-1}$ is reached at the end of the main sequence for the 3\,M$_\odot$ star presented here. A very rapid decrease of the surface velocity then occurs when the star crosses the Hertzsprung gap. The exact value of the surface velocity during the red giant phase is sensitive to the assumption made on the rotation law in the extended convective envelope of the star. By assuming solid body rotation or uniform specific angular momentum in the external convective zone, a mean value of the surface rotational velocity of about 6 and 0.3\,km\,s$^{-1}$ is respectively found during the red giant phase.     
For both limiting assumptions on the rotation law in the external convective zone, a model computed with a significant value for the initial rotational velocity on the ZAMS thus exhibits a low surface rotational velocity as a red giant. This shows that a slowly rotating red giant may have been rotating much more rapidly during the main sequence and that its global properties depend on its rotational history.

\subsection{Low-mass stars}
\label{sec_dhr_m1}

\begin{figure}[htb!]
\includegraphics[scale=.55]{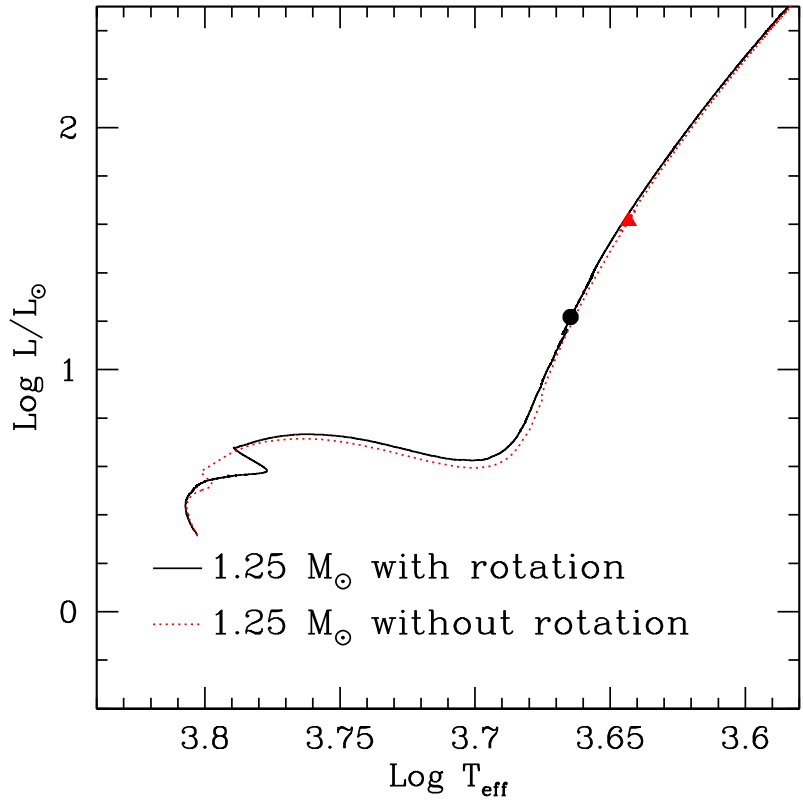}
\caption{Evolutionary tracks in the HR diagram for 1.25\,M$_\odot$ models computed with (continuous line) and without (dotted line) shellular rotation. The dot and triangle indicate the location of the bump for the rotating and non-rotating model, respectively.}
\label{dhr_m1p25}
\end{figure}

The impact of rotational mixing on low-mass red giants that undergo the helium flash at the RGB tip is now briefly discussed by comparing 1.25\,M$_\odot$ models with and without rotation. These models are computed with the evolution code STAREVOL \citep{cha10} at solar metallicity, with an initial velocity on the ZAMS of 110\,km\,s$^{-1}$ for the rotating case. The corresponding evolutionary tracks are shown in Fig.~\ref{dhr_m1p25}. 

During the evolution on the main sequence, the inclusion of rotational effects results in slightly larger effective temperatures and luminosities. For these models with a lower mass and a small convective core, this shift is mainly due to the fact that rotational mixing counteracts the effects of atomic diffusion in the external layers of the star \citep{egg10_sl}. Larger values of helium abundance are then found at the surface of models including rotation, which leads to a decrease of the opacity in the external layers of the star. As for the more massive 3\,M$_\odot$ models described above, the influence of rotational mixing is not restricted to the external layers of the star, since the properties of the central layers are changed due to the transport of fresh hydrogen 
fuel into the stellar core. Consequently, the main-sequence lifetime is larger when rotation is accounted for.
Contrary to the case of intermediate-mass red giants discussed in Sect.~\ref{sec_dhr_m3}, the location in the HR diagram of the core helium burning phase is now very similar for models computed with and without rotational effects. However, rotational mixing has a large impact on the location of the bump. At solar metallicity, the inclusion of rotation leads indeed to a significant decrease of the luminosity of the star at the bump and to an increase of its effective temperature \citep[see Fig.~\ref{dhr_m1p25} and][]{cha10}.

\section{Effects of rotation on asteroseismic properties of red giants}
\label{sec_astero}

Red giant stars are characterised by deep convective envelopes and dense cores. 
From an asteroseismic point of view, the large density near the center of the star leads to huge values 
of the Brunt-V\"ais\"al\"a frequency in the central layers. Consequently, oscillation modes of mixed p-mode and g-mode 
properties are expected. In addition to purely acoustic radial modes, 
a large number of non-radial modes are then found, which are in most cases dominated by the g-mode behaviour 
and have therefore a high inertia. However, for non-radial modes trapped 
in the acoustic cavity the inertia becomes similar to the one of radial modes. For $\ell=2$ modes, 
the separation between the p- and g-mode region is sufficient to obtain oscillation modes 
which in terms of inertia are very similar to purely acoustic modes. This results in 
trapped $\ell=2$ modes with a dominant amplitude close to every radial mode, which define 
the small separation between $\ell=0$ and $\ell=2$ modes \cite[see e.g.][]{chr04b}. 
This basic description of the properties of oscillation modes in red giant stars is solely based on inertia consideration 
and does not take into account the problem of excitation and damping of these modes \citep[see e.g.][]{dzi01, dup09}. 

\subsection{Intermediate-mass stars}
\label{sec_astero_m3}

We now discuss the effects of rotation on the properties of the oscillation modes of these
stars. This is first done by comparing the asteroseismic properties of rotating and non-rotating models at the same evolutionary phase
computed with the same input parameters (except for the inclusion of shellular rotation).
A rotating model of 3\,M$_\odot$ computed with an initial velocity of 150\,km\,s$^{-1}$ on the ZAMS and situated in the middle of the core helium-burning phase (Y$_{\rm c}=0.355$) is then compared to the corresponding non-rotating red-giant model. The theoretical low-$\ell$ frequencies of both models are computed 
and the values of the large and small frequency separations are obtained. The values of the large separation are calculated from the radial modes, while the small separations between $\ell=0$ and $\ell=2$ modes are determined by considering only $\ell=2$ modes well trapped in the acoustic cavity. 

Rotational effects lead to a significantly lower value of the mean large separation. The rotating model exhibits indeed a mean large separation that is about  20\% lower that the one of the non-rotating model. The mean large separation being mainly proportional to the square root of the star's mean density, this difference is directly related to the larger radius of the rotating model (both models share the same mass of 3\,M$_\odot$). This larger radius is a direct consequence of the significant increase of the luminosity discussed in the preceding sections when rotational mixing is included in the computation. The value of the mean small frequency separation between radial modes and $\ell=2$ modes trapped in the acoustic cavity of the star is also found to be significantly reduced when rotational effects are taken into account. For stars evolving on the main sequence, the ratio of the small to large separation is sensitive to the conditions in the central regions of the star \citep{rox03,rox05}. In the case of red giant models, 
the decrease of the values of the small and large separation when rotation is included is similar, 
so that the ratio between the small and large separation remains approximately 
the same for rotating and non-rotating models. This shows that the decrease of the small separation between $\ell=0$ and trapped $\ell=2$ modes observed for rotating models is mainly due to the change of the global stellar properties 
and not to a change in the structure of the central regions of the star \citep[][]{egg10_rg,mon10}.   

In addition to the comparison between rotating and non-rotating models computed with the same initial parameters, it is also interesting to discuss the impact of rotation on the determination of the fundamental stellar parameters and asteroseismic properties for red giants sharing 
the same location in the HR diagram. This is done by computing another model with approximately 
the same luminosity as the non-rotating red giant stars of 3\,M$_{\odot}$ during the core helium-burning phase. Since rotational mixing increases the stellar luminosity, such a rotating model is obtained for a lower mass of  2.7\,M$_{\odot}$. The lower initial mass of models including rotation leads to a large increase of the age determined for a red giant. This illustrates how rotational mixing changes the global stellar parameters needed to reach the same location in the HR diagram for core helium-burning stars.   

The change of the global stellar parameters induced by rotation also results in differences in the asteroseismic properties between rotating and non-rotating models of red giants sharing the same location in the HR diagram.
The rotating 2.7\,M$_{\odot}$ model is then characterised by  
a lower value of the mean large separation than the 3\,M$_{\odot}$ non-rotating model. This directly reflects the different masses of both models, which lead to different mean densities and hence mean large frequency separations. The mean small frequency separation between radial and $\ell=2$ modes trapped in the acoustic cavity of the star is also found to decrease when rotational effects are taken into account. As mentioned above, 
this is mainly due to the change of the global stellar properties (the stellar mass in this case) for models computed with rotation compared to non-rotating models and not to differences in the structure of the central stellar layers.

\subsection{Low-mass stars}
\label{sec_astero_m1}

The global asteroseismic properties of a rotating (initial velocity of 110\,km\,s$^{-1}$ on the ZAMS) 1.25\,M$_\odot$ red-giant model at the bump and the corresponding non-rotating model situated at the same evolutionary stage are finally compared. As noted above, the values of the large separation are calculated from the radial modes, while the small separations between $\ell=0$ and $\ell=2$ modes are determined by considering only $\ell=2$ modes well trapped in the acoustic cavity of the star. For low-mass red giants at the bump, the inclusion of rotation results in a large increase of the mean value of the large separation, which is directly related to the smaller radius of the rotating model. As seen in Sect.~\ref{sec_dhr_m1}, a rotating model of 1.25\,M$_\odot$ at solar metallicity is characterised by a lower luminosity and a larger effective temperature at the bump than a model without rotation. This leads to a smaller radius and hence a larger mean density and large frequency separation when rotational effects are taken into account. As for more massive red giant models, this change of the  global stellar properties also results in a change of the mean small frequency separation between radial modes and trapped $\ell=2$ modes with a significant increase of the small separation for rotating red giants.

\section{Summary}

The surface rotation velocity of a red giant star is generally low, but the star may have been rotating much more rapidly during the main sequence. 
The evolution in the red giant phase is then sensitive to the rotational history of the star, because rotation significantly changes its internal structure and global properties during the main sequence. Rotational mixing is the main driver of the changes of the evolutionary tracks in the HR diagram, with only a very limited contribution from hydrostatic corrections induced by rotation. For red giants massive enough to ignite He burning in non-degenerate conditions, rotational mixing shifts the location of the core helium burning phase to higher luminosity in the HR diagram, while for low-mass red giants undergoing the helium flash the luminosity at the bump is significantly reduced when rotation is included in the computation. This of course results in different global asteroseismic properties for red giant models with and without rotation and changes the values of the fundamental stellar parameters (in particular the mass and age of the star) determined from an asteroseismic calibration. In addition to these effects of rotation on the global asteroseismic properties, it will be interesting to study in detail the effects of rotational mixing on asteroseismic observables that are more sensitive to changes in the structure near the central core like the small separation between radial and dipole modes \citep{mon10}.
Asteroseismic data coming from ground-based observations and space missions are thus particularly valuable to provide us with new insight into transport processes in stellar interiors like rotation, magnetic fields \citep[e.g.][]{egg10_magn}, and internal gravity waves \citep[e.g.][]{cha05}.

\begin{acknowledgement}
Part of this work is supported by the Swiss National Science Foundation and by the French Programme National de Physique Stellaire (PNPS) of CNRS/INSU.
\end{acknowledgement}

\end{document}